\documentclass[aps,pra,showpacs,twocolumn]{revtex4-1}%
\usepackage{amsfonts}
\usepackage{amsmath}
\usepackage{amssymb}
\usepackage{graphicx}%
\providecommand{\U}[1]{\protect\rule{.1in}{.1in}}

\begin{document}
\title{Simulating and Detecting the Quantum Spin Hall Effect in Kagom\'{e} Optical Lattice}
\author{Guocai Liu$^{1}$, Shi-Liang Zhu$^{2}$, Shaojian Jiang$^{1}$, Fadi Sun$^{1}$,
W. M. Liu$^{1}$}
\affiliation{$^{1}$Beijing National Laboratory for Condensed Matter Physics, Institute of
Physics, Chinese Academy of Science, Beijing 100190, China }
\affiliation{$^{2}$Laboratory of Quantum Information Technology, ICMP and SPTE, South China
Normal University, Guangzhou, China}


\begin{abstract}
We propose a model which includes a nearest-neighbor intrinsic spin-orbit
coupling and a trimerized Hamiltonian in the Kagom\'{e} lattice and promises
to host the transition from the quantum spin Hall insulator to the normal
insulator. In addition, we design an experimental scheme to simulate and
detect this transition in the ultracold atom system. The lattice intrinsic
spin-orbit coupling is generated via the laser-induced-gauge-field method.
Furthermore, we establish the connection between the spin Chern number and the
spin-atomic density which enables us to detect the quantum spin Hall insulator
directly by the standard density-profile technique used in the atomic systems.

\end{abstract}
\maketitle

\section{INTRODUCTION}

Optical lattice system has gradually become a promising platform to simulate
and study a lot of quantum phenomena in condensed matter physics because
almost all parameters of the system can be well-controlled\cite{Lewe}. The
recent theoretical and experimental progress in laser-induced-gauge-field
\cite{Ruse,Oste,Juze,Zhu2,Lin1,Lin2} makes it a hot spot to study topological
quantum states \cite{Klit} in cold atoms
system\cite{Stanescu1,Bermudez,Satija,Zhu3,LXJ2,Gold2,Varn}. Subject to the
compounds' natural properties \cite{Stan}, the famous Haldane model
\cite{Hald} proposed two decades ago has not been confirmed by experiments
because the required periodic magnetic field cannot be easily implemented in
actual material. For topological insulator, an important extension from the
Haldane model to a time-reversal invariant system
\cite{Kane,Bern,Fu,Fu2,Moore2,Butt}, only a few materials are confirmed
currently to have such exotic topological properties in the nature
\cite{Mura,Koni,Hsieh3,Zhang}, because the existence of such properties
require relatively strong spin-orbit (SO) coupling. However, in optical
lattice system, we can engineer the lattice Hamiltonian to guarantee that this
system hosts these novel topological phases \cite{Gold2}. Up to now, the
neutral-cold-atom integer and fractional quantum Hall effects have been
studied \cite{Lukin,Palm,Umuc} and also the realization of Haldane model has
been designed by using the laser-induced-gauge-field method in optical lattice
\cite{Zhu3}.

In this paper, we propose a scheme to simulate and detect the 2-dimensional
(2D) quantum spin Hall (QSH) insulator in a Kagom\'{e} optical lattice with a
trimer and a nearest-neighbor SO coupling term. With laser-induced-gauge-field
method, one can design a variety of lattice SO couplings
\cite{Zhu2,XJLiu,Li,Stanescu}, which is convenient for us to study the 2D
topological insulator in the optical lattice \cite{Stanescu1,Bermudez,Satija}.
However, the original proposal \cite{Kane} of realizing 2D QSH insulator in
honeycomb lattice requires the next-nearest-neighbor hopping amplitude. This
requirement causes doing experiments in the optical lattice difficult because
high barrier makes the next-nearest-neighbor tunneling very small. Recently, a
model \cite{Guo} raised in Kagom\'{e} lattice also requires the
next-nearest-neighbor hopping. Interestingly, there is another QSH insulator
model in the complicated Kagom\'{e} lattice. The inspiration comes from the
fact that spin chirality in ferromagnetic Kagom\'{e} lattice exerts important
effects on orbital magnetic moment and anomaly quantum Hall effect
\cite{Wang2}. We find that a Kagom\'{e} optical lattice with the trimer and SO
coupling terms can host the 2D QSH insulator phase with only the
nearest-neighbor hopping. Since the model only involves the nearest-neighbor
hopping, it would be easier to be implemented in cold atomic experiments.

Furthermore, compared to the condensed matter systems, we find that detecting
QSH insulator has more advantage in optical lattice system. Due to the
time-reversal symmetry, Chern number can not be taken as a topological
invariant to characterize the QSH insulator in real electron system
\cite{Sheng}. Chern number $C_{\uparrow}$=$+1$ ($C_{\downarrow}$=$-1$) in a
QSH phase for the up-spin (down-spin) electrons and the total Chern number
$C$=$C_{\uparrow}$+$C_{\downarrow}$=$0$. Since one can not distinguish the
contributions of the conductance from the up or down spin electrons at current
technology, it is impossible to determine whether the system is in the QSH
phase or normal phase by measuring Hall conductance. However, in cold atomic
systems, it is the atom's internal states that represent the spin, not real
spin, which brings certain benefits to measurement. One can directly measure
the spin Chern number to determine whether the system lies in the QSH phase or
not because optically measuring the atomic internal states is very simple. In
this paper, we demonstrate that the method developed to detect the Chern
number in cold atomic systems \cite{Zhu3,Umuc} can be put forward further to
measure the spin Chern number, and thus we establish the connection between
the spin Chern number and the spin-atomic density which enables us to detect
the topological Chern numbers directly by the standard density-profile
technique used in the atomic systems.

The paper is organized as follows: In Sec. II, we introduce the model in
Kagom\'{e} lattice with both trimer and spin-orbital coupling. This model can
realize the QSH phase with only the nearest-neighbor hoping terms; in Sec.
III, we explain how to simulate this model in cold-atom optical lattice, which
includes designing the lattice SO coupling by using the
laser-induced-gauge-field method; we present the method to detect the QSH
phase in Sec. IV and give a brief summary in Sec. V.

\section{MODEL}

Let us consider the tight-binding model for two-component fermionic atoms on
the Kagom\'{e} optical lattice, which consists of three triangular sublattices
A, B and C (Fig. 1). The spin-independent part of the Hamiltonian is given by%
\begin{align}
H_{0}  &  \text{=}t_{0}\sum_{mn\alpha}\left(  b_{m,n,\alpha}^{\dag
}a_{m,n,\alpha}\text{+}b_{m-1,n,\alpha}^{\dag}a_{m,n,\alpha}\right.
\nonumber\\
&  \left.  \text{+}c_{m,n,\alpha}^{\dag}b_{m,n,\alpha}\text{+}%
c_{m+1,n-1,\alpha}^{\dag}b_{m,n,\alpha}\right. \nonumber\\
&  \left.  \text{+}a_{m,n,\alpha}^{\dag}c_{m,n,\alpha}\text{+}a_{m,n+1,\alpha
}^{\dag}c_{m,n,\alpha}\right)  \text{+H.c.}, \label{Hnn}%
\end{align}
where $t_{0}$ is the hopping amplitude between the nearest neighbor link,
$(m,n)$ labels the Kagom\'{e} unit cells with the unit vectors $\mathbf{b}%
_{1}$=$\left(  2,0\right)  a$ and $\mathbf{b}_{2}$=$\left(  1,\sqrt{3}\right)
a$, $a_{m,n,\alpha}^{\dag}$ ($a_{m,n,\alpha}$) is the creation (annihilation)
operator of an atom with spin $\alpha$ (up or down) on lattice site $(m$,$n)$
on sublattice A (an equivalent definition is used for sublattice B and C). For
simplicity, we choose $t_{0}$=$1$ as the energy unit and the distance between
the nearest sites $a$ as the length unit throughout this
paper.\begin{figure}[ptb]
\begin{center}
\includegraphics[width=1.0\linewidth]{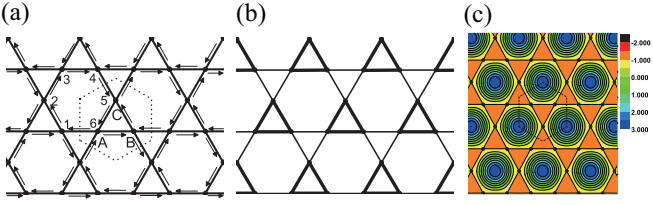}
\end{center}
\caption{(Color online) (a) Schematic picture of the nearest neighbor
intrinsic SO coupling in 2D Kagom\'{e} lattice. The up-spin atoms hop along
(against) the arrowed direction with amplitude $i\lambda_{\text{SO}}$
($-i\lambda_{\text{SO}}$). For the down-spin atoms, the arrows are reversed.
The dashed line represents the Wigner-Seitz unit cell, which contains three
independent sites (A, B, C). (b) the trimer Kagom\'{e} lattice. Hopping
amplitude corresponds to $t\mathtt{+}\kappa$ ($t\mathtt{-}\kappa$) for the
thick (thin) bonds. (c) Contours of the effective magnetic field for up-spin
atoms defined by Eq. (\ref{B1}).}%
\end{figure}

By using the Fourier transform of atomic operators $a_{m,n,\alpha}$, i.e.,
\begin{equation}
a_{m,n,\alpha}=\frac{1}{\sqrt{N}}\sum_{\mathbf{k}}a_{\mathbf{k}\alpha
}e^{-i\mathbf{k\cdot R}_{mn}^{A}} \label{Fourier}%
\end{equation}
the Hamiltonian (\ref{Hnn}) can be diagonalized in the momentum space as%
\begin{equation}
H_{0}=\sum_{\mathbf{k}}\psi_{\mathbf{k}}^{+}(\mathcal{H}_{0}(\mathbf{k}%
)\mathtt{\otimes}\mathbf{I}_{2\times2})\psi_{\mathbf{k}}, \label{Hk}%
\end{equation}
where $\mathbf{I}_{2\times2}$ is a $2\mathtt{\times}2$ unit matrix. The
six-component basis operator $\psi_{\mathbf{k}}$=$(a_{\mathbf{k}\uparrow
},b_{\mathbf{k}\uparrow},c_{\mathbf{k}\uparrow},a_{\mathbf{k}\downarrow
},b_{\mathbf{k}\downarrow},c_{\mathbf{k}\downarrow})^{\text{T}}$ describes an
atom in the corners of the Kagom\'{e} lattice unit cell (Fig. 1) with spin
$\alpha$ ($\uparrow$ and $\downarrow$). Here $\mathcal{H}_{0}(\mathbf{k})$ is
a $3\mathtt{\times}3$ spinless matrix given by%
\begin{equation}
\mathcal{H}_{0}(\mathbf{k})\text{=}\left(
\begin{array}
[c]{ccc}%
0 & 2\cos P_{1} & 2\cos P_{3}\\
2\cos P_{1} & 0 & 2\cos P_{2}\\
2\cos P_{3} & 2\cos P_{2} & 0
\end{array}
\right)  , \label{Hamilton}%
\end{equation}
where we have defined $P_{1}$=$\mathbf{k}\mathtt{\cdot}\mathbf{a}_{1}$,
$P_{2}$=$\mathbf{k}\mathtt{\cdot}\mathbf{a}_{2}$, $P_{3}$=$\mathbf{k}%
\mathtt{\cdot}\mathbf{a}_{3}$ with $\mathbf{a}_{1}$=$(1,0)$, $\mathbf{a}_{2}%
$=$(-1/2,\sqrt{3}/2)$, and $\mathbf{a}_{3}$=$(-1/2,-\sqrt{3}/2)$ representing
the displacements in a unit cell from A to B site, from B to C site, and from
C to A site, respectively. In this notation, the first Brillouin zone is a
hexagon with the corners of $\mathbf{K}$=$\pm\left(  2\pi/3\right)
\mathbf{a}_{1}$, $\pm\left(  2\pi/3\right)  \mathbf{a}_{2}$, $\pm\left(
2\pi/3\right)  \mathbf{a}_{3}$, and two of which are independent.

The energy spectrum for spinless Hamiltonian $\mathcal{H}_{0}(\mathbf{k})$ is
characterized by one dispersionless flat band ($\epsilon_{1\mathbf{k}}^{(0)}%
$=$-2$), which reflects the fact that the 2D Kagom\'{e} lattice is a line
graph of the honeycomb structure \cite{Mie}, and two dispersive bands,
$\epsilon_{2(3)\mathbf{k}}^{(0)}$=$1\mathtt{\mp}\sqrt{4b_{\mathbf{k}}-3}$ with
$b_{\mathbf{k}}$=$\sum_{i=1}^{3}\cos^{2}\left(  \mathbf{k\cdot a}_{i}\right)
$. These two dispersive bands touch at Dirac points $\mathbf{K}$ and exhibit a
cusp, $\epsilon_{2(3)\mathbf{k}}^{(0)}\mathtt{=}(1\mathtt{\mp}\sqrt
{3}|\mathbf{k}\mathtt{-}\mathbf{K}|)$.

When the following intrinsic SO coupling term is taken into account in the
Kagom\'{e} lattice model, as illustrated in Fig. 1(a), the gap will be opened
at the two inequivalent Dirac points. The tight-binding expression for this SO
coupling Hamiltonian can be given as follows:%
\begin{align}
H_{\text{SO}}  &  \text{=}i\lambda_{\text{SO}}\sum_{m,n}\left(  b_{m,n}^{\dag
}\sigma_{z}a_{m,n}\text{+}b_{m-1,n}^{\dag}\sigma_{z}a_{m,n}\right. \nonumber\\
&  \left.  \text{+}c_{m,n}^{\dag}\sigma_{z}b_{m,n}\text{+}c_{m+1,n-1}^{\dag
}\sigma_{z}b_{m,n}\right. \nonumber\\
&  \left.  \text{+}a_{m,n}^{\dag}\sigma_{z}c_{m,n}\text{+}a_{m,n+1}^{\dag
}\sigma_{z}c_{m,n}\right)  \text{+H.c.}, \label{HSO}%
\end{align}
where $\lambda_{\text{SO}}$ is the SO coupling constant,\textbf{\ }$\sigma
_{z}$ is the Pauli matrix and $a_{m,n}^{\dag}$=$(a_{m,n,\uparrow}^{\dag
},a_{m,n,\downarrow}^{\dag})$. Taking the Fourier transform (\ref{Fourier})
and considering the $\psi_{\mathbf{k}}$ below Eq. (\ref{Hk}), we have
$H_{\text{SO}}\mathtt{=}\sum_{\mathbf{k}}\psi_{\mathbf{k}}^{+}\mathcal{H}%
_{\text{SO}}(\mathbf{k})\psi_{\mathbf{k}}$, where
\begin{equation}
\mathcal{H}_{\text{SO}}(\mathbf{k})=\left(
\begin{array}
[c]{cc}%
\mathcal{H}_{+}(\mathbf{k}) & 0\\
0 & \mathcal{H}_{-}(\mathbf{k})
\end{array}
\right)
\end{equation}
with%
\begin{equation}
\mathcal{H}_{\pm}(\mathbf{k})\text{=}\mathtt{\pm}2i\lambda_{\text{SO}}\left(
\begin{array}
[c]{ccc}%
0 & \mathtt{-}\cos P_{1} & \cos P_{3}\\
\cos P_{1} & 0 & \mathtt{-}\cos P_{2}\\
\mathtt{-}\cos P_{3} & \cos P_{2} & 0
\end{array}
\right)  . \label{HR}%
\end{equation}
This SO coupling destroys spin SU(2) symmetry and opens a band gap
$\Delta_{\text{SO}}$=$\sqrt{3}\lambda_{\text{SO}}$ at Dirac point.

Lattice trimerization can break inversion symmetry of Kagom\'{e} lattice and
also open a gap at Dirac point \cite{Guo}. It is described by%
\begin{align}
H_{\text{trim}}  &  \text{=}\sum_{mn\alpha}\left[  \kappa\left(
b_{m,n,\alpha}^{\dag}a_{m,n,\alpha}-b_{m-1,n,\alpha}^{\dag}a_{m,n,\alpha
}\right)  \right. \nonumber\\
&  \left.  +\kappa\left(  c_{m,n,\alpha}^{\dag}b_{m,n,\alpha}%
-c_{m+1,n-1,\alpha}^{\dag}b_{m,n,\alpha}\right)  \right. \nonumber\\
&  \left.  +\kappa\left(  a_{m,n,\alpha}^{\dag}c_{m,n,\alpha}-a_{m,n+1,\alpha
}^{\dag}c_{m,n,\alpha}\right)  \right]  +\text{H.c}, \label{HD}%
\end{align}
where $\kappa$ describe an alternating pattern of bond hopping integrals along
the three principal spatial directions as illustrated in Fig. 1(b). Taking the
Fourier transform (\ref{Fourier}) again, the trimerized Hamiltonian can be
rewritten as $H_{\text{trim}}\mathtt{=}\sum_{\mathbf{k}}\psi_{\mathbf{k}}%
^{+}(\mathcal{H}_{\text{trim}}(\mathbf{k})\mathtt{\otimes}\mathbf{I}%
_{2\times2})\psi_{\mathbf{k}}$ with%
\begin{equation}
\mathcal{H}_{\text{trim}}\left(  \mathbf{k}\right)  \text{=}2i\left(
\begin{array}
[c]{ccc}%
0 & -\kappa\sin P_{1} & \kappa\sin P_{3}\\
\kappa\sin P_{1} & 0 & -\kappa\sin P_{2}\\
-\kappa\sin P_{3} & \kappa\sin P_{2} & 0
\end{array}
\right)  \label{HDK}%
\end{equation}
for both spin components.

We take above SO coupling and trimerized Hamiltonian as perturbation, which
means $\lambda_{\text{SO}}\mathtt{\ll}t$ and $\kappa\mathtt{\ll}t$. Although
both of perturbations can bring gaps at Dirac points independently, these two
gaps have different topological nature. As we will show in the following, the
former is non-trivial and quantum spin Hall effect will occur if Fermi energy
level locates in the gap; the latter is a trivial gap.

In order to prove above assertion, we expand the total Hamiltonian
$\mathcal{H}\left(  \mathbf{k}\right)  $=$\mathcal{H}_{0}\left(
\mathbf{k}\right)  $+$\mathcal{H}_{\text{SO}}\left(  \mathbf{k}\right)
$+$\mathcal{H}_{\text{trim}}\left(  \mathbf{k}\right)  $ at two inequivalent
points $\mathbf{K}_{\pm}$=$\left(  \pm\frac{2\pi}{3},0\right)  $, then take
$\mathbf{k\cdot p}$ perturbation theory to get its effective Hamiltonian. At
last, projecting it onto bands 2 and 3 subspace, we get four independent Dirac
Hamiltonian,%
\begin{equation}
\mathcal{H}_{s\sigma}^{K}\text{=}-s\upsilon_{F}k_{x}\tau_{z}+s\upsilon
_{F}k_{y}\tau_{x}+m_{s\sigma}\tau_{y} \label{HK1}%
\end{equation}
where $s$=$\pm1$ and $\sigma$=$\pm1$ represent different valleys
$\mathbf{K}_{\pm}$ and spin indices, respectively, $\upsilon_{F}$=$\sqrt{3}t$
is Fermi velocity, and $\tau_{i}$ are Pauli matrices with $i$=$x,y,z$. The
Dirac mass $m_{s\sigma}$=$\sqrt{3}\sigma\lambda_{\text{SO}}\mathtt{-}3s\kappa
$.\begin{figure}[ptb]
\begin{center}
\includegraphics[width=0.6\linewidth]{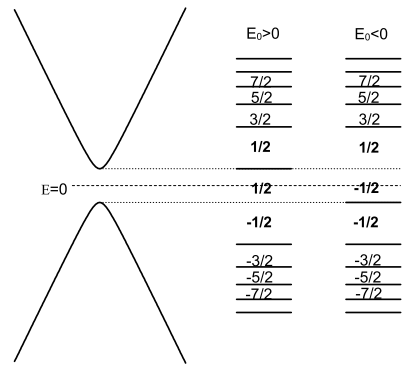}
\end{center}
\caption{(Color online) Illustration of the effect on Chern number by the sign
of $n$=$0$ Landau energy level. The left figure corresponds to disperse
relation of Eq. (\ref{HK1}) without magnetic field. The right two figure show
the Chern numbers when magnetic field is applied and Fermi energy locate at
different interval for two cases with $E_{0}>0$ and $E_{0}<0$. Here $E_{0}$
denotes the $n$=$0$\ Landau energy level. The dashed line represents the zero
energy and the doted lines give the original band gap. }%
\end{figure}

We now turn to address the topological properties of this model. The system is
in the QSH phase when the SO coupling dominates in the condition $\left\vert
\lambda_{\text{SO}}\right\vert \mathtt{>}\left\vert \sqrt{3}\kappa\right\vert
$; otherwise, the system is in the normal phase. It can be proved by directly
calculating the $Z_{2}$ topological invariant \cite{Kane,Fu}. Alternatively,
one can also understand the topological phases from the view of spin Chern
number \cite{Sheng}. In order to do that, we apply a uniform magnetic field
along $z$-direction with gauge vector $\mathbf{A}\left(  \mathbf{r}\right)
=\left(  0,\mathcal{B}x,0\right)  $ and obtain the Landau energy levels%
\begin{equation}
E_{n}^{s\sigma}=\left\{
\begin{array}
[c]{l}%
m_{s\sigma}\text{sgn}\left(  e\mathcal{B}\right) \\
\multicolumn{1}{c}{\pm\sqrt{2n\hbar\upsilon_{F}^{2}\left\vert e\mathcal{B}%
\right\vert +m_{s\sigma}^{2}}}%
\end{array}%
\begin{array}
[c]{l}%
n=0\\
\multicolumn{1}{c}{n=1,2,3,\cdots}%
\end{array}
\right.  \label{LE}%
\end{equation}
By using the Green function theory \cite{Zhu3}, we can get the Chern number
\begin{equation}
C_{s\sigma}=\frac{1}{2}\text{sgn}(E_{0}^{s\sigma})=\frac{1}{2}\text{sgn}%
(m_{s\sigma}\text{sgn}\left(  e\mathcal{B}\right)  )
\end{equation}
between the energy interval $-\left\vert m_{s\sigma}\right\vert \mathtt{<}%
\mu\mathtt{<}\left\vert m_{s\sigma}\right\vert $ with $\mu$ being the Fermi
energy (Fig. 2). Actually it is the sign of $n$=$0$ energy (zero-mode) which
determines the Chern number. As an example, We focus on $\mu\mathtt{=}0$ and
take $e\mathcal{B}\mathtt{>}0$ throughout this paper to see the difference
more clearly between QSH phase and normal phase. For QSH phase, we take
$\lambda_{\text{SO}}\mathtt{>}\sqrt{3}\kappa\mathtt{>}0$, and it is easy to
see that, for up-spin atoms, $m_{s=\pm1,\uparrow}\mathtt{>}0$ and then
C$_{\uparrow}$=C$_{+1,\uparrow}\mathtt{+}C_{-1,\uparrow}$=$1$. For down-spin
atom, $m_{s=\pm1,\downarrow}\mathtt{<}0$ and C$_{\downarrow}$%
=C$_{+1,\downarrow}\mathtt{+}C_{-1,\downarrow}$=$-1$. However, if we take
$\sqrt{3}\kappa\mathtt{>}\lambda_{\text{SO}}\mathtt{>}0$, which corresponds to
the normal phase, C$_{\uparrow}$=C$_{\downarrow}$=$0$ for both up- and
down-spin atoms. In the QSH phase, the Chern numbers with different spin
components have same value but with opposite sign. Whereas if it is in normal
phase, the Chern numbers equal to zero for both up- and down-spin atoms.

To further understanding the topological properties of the model, we show the
edge state effects in Fig. 3. From Fig. 3(a), we can see that there is a pair
of chiral gapless edge states for every band gap when the SO coupling
dominates. This means that the system is in topological insulator phases at
1/3- and 2/3-filling. When only trimer term exists, it opens a band gap at
Dirac point but no edge states connect the upper and lower bands (Seeing Fig.
3(b)), therefore the system is in normal insulator phase at 2/3-filling. On
the other hand we also see that the trimer term cannot open a gap between the
band 1 and 2 (Fig. 3(b)). Therefore the system at 1/3-filling will be still in
the topological insulator phase when two perturbations are present but the
trimer term dominates (Seeing Fig. 3(c)). \begin{figure}[ptb]
\begin{center}
\includegraphics[width=1.0\linewidth]{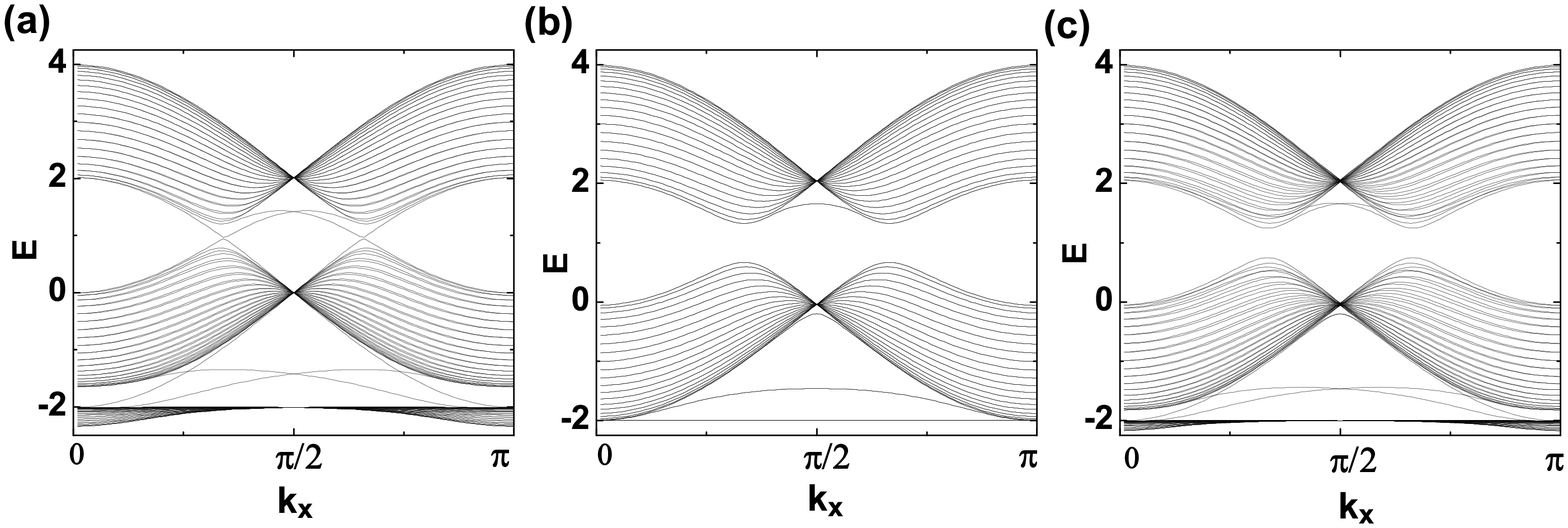}
\end{center}
\caption{(Color online) The band structure of the lattice model in the stripe
geometry. We take $\lambda_{\text{SO}}$=$0.1$, $\kappa$=$0$ for (a),
$\lambda_{\text{SO}}$=$0$, $\kappa$=$0.1$ for (b) and $\lambda_{\text{SO}}%
$=$0.05$, $\kappa$=$0.1$ for (c).}%
\end{figure}

\section{Simulation}

In this section we introduce an approach to simulate the Kagom\'{e} lattice
with the trimer and SO coupling terms in an optical lattice system. To this
end, two problems need to be solved. One is how to generate the Kagom\'{e}
optical lattice with the trimer terms; the other is how to simulate the
lattice SO coupling. As for the first problem, we use the superlattice
technique addressed in Refs. \cite{Sant,Dams,Lee}, that is, three super-laser
beams with the same wave vector length but different polarizations are applied
along three different directions: $\mathbf{e}_{y}$ and $\frac{1}{2}%
\mathbf{e}_{y}\mathtt{\pm}\frac{\sqrt{3}}{2}\mathbf{e}_{x}$, respectively. In
our proposal, each super-laser beam consists of four large detuned
standing-wave lasers with the same polarization but different wave vector
length in the $x$-$y$ plane. The total potential is thus given by%
\begin{align}
V\left(  \mathbf{r}\right)   &  \text{=}V_{0}\sum_{i=1}^{3}\left[  \cos\left(
\mathbf{k}_{i}\mathbf{\cdot r}\text{+}3\delta_{i}\varphi/2\right)
\text{+}2\cos\left(  \mathbf{k}_{i}\mathbf{\cdot r}/3\text{+}\delta_{i}%
\varphi/2\right)  \right. \nonumber\\
&  \left.  \text{+}4\cos\left(  \mathbf{k}_{i}\mathbf{\cdot r}/9\text{+}%
\delta_{i}\varphi/6\right)  \right. \nonumber\\
&  \left.  \text{+}\zeta\cos\left(  \mathbf{k}_{i}\mathbf{\cdot r}%
/9\text{+}\delta_{i}\left(  \varphi/6\text{+}\pi/2\right)  \right)  \right]
^{2} \label{V1}%
\end{align}
with the wave vectors $\mathbf{k}_{1}$=$(\frac{\sqrt{3}}{2},\frac{1}{2})k$,
$\mathbf{k}_{2}$=$(-\frac{\sqrt{3}}{2},\frac{1}{2})k$, $\mathbf{k}_{3}%
$=$\left(  0,1\right)  k$ and $\delta_{1}$=$\delta_{2}$=$-\delta_{3}$=$1$.
Firstly, we consider the case with $\zeta$=$0$. One can get a triangular
lattice when $\varphi$=$0$ or $2\pi$ and a Kagom\'{e} lattice when
$0\mathtt{<}\varphi\mathtt{<}2\pi$. A uniform Kagom\'{e} lattice corresponds
to $\varphi$=$\pi$, as shown in Fig. 4(a). When $\varphi$ takes other values,
one can obtain the trimerized Kagom\'{e} lattice accompanying distortion of
the lattice structure \cite{Sant,Dams,Lee}. When we increase the strength of
trimerized Hamiltonian, the Kagom\'{e} lattice will be distorted. To overcome
this defect, we add another laser beam which corresponds to the $\zeta
\mathtt{\neq}0$ in the Eq. (\ref{V1}) and assume $\varphi$=$\pi$ all the time.
The added laser will interfere with primary lasers and generate the trimerized
Kagom\'{e} lattice. $\zeta$ is an adjustable parameter to control the strength
of trimerized Hamiltonian. With this method, the Kagome lattice will not have
obvious offset from the uniform one even if $\zeta$ takes a relative large
value. As shown in Fig. 4(b), the parameter $\zeta$=$1.5$ is chosen as a
typical example.

We now focus on how to simulate the lattice SO coupling. Using the
laser-induced-gauge-field method, it was proposed that both Abelian and
non-Abelian gauge fields can be simulated in cold atomic system. In addition,
the experiments to achieve such artificial gauge fields have been reported
\cite{Lin1,Lin2}. Interestingly, it was proposed that a periodic magnetic
field, which is not easy to be realized in a condensed-matter system, can be
created by two opposite-traveling standing-wave laser beams \cite{Zhu2}. In
the following, we will show in an explicit manner that the artificial gauge
field proposed in Ref. \cite{Zhu2} is equivalent to a SO coupling.
Furthermore, the required lattice SO coupling addressed in the previous
section can also be achieved in a suitable configuration of the laser beams.

In optical lattice systems, cold atoms can hop between adjacent sites.
According to Peierls theory, the additional gauge vector potential
$\mathbf{A}$ makes the hopping obtain a phase factor $\exp\left(  i\frac
{e}{\hbar}\int\mathbf{A\cdot dl}\right)  $, where the integral along the
hopping path. If the atom has multiple states, which correspond to different
spin components, vector potential $\mathbf{A}$ should be a matrix. After
taking such an approximation, that is $t_{ij}^{\alpha\beta}$=$t_{ij}$, here
$\alpha,\beta$ on behalf of any spin index, the correction to atom hopping
between different sites coming from gauge field is equivalent to a unitary
operator \cite{Gold}. For a two-component atom system, the unitary operator
can be written as%
\begin{equation}
U_{ij}=e^{i\alpha_{ij}\sigma_{ij}}=\cos\alpha_{ij}+i\sigma_{ij}\sin\alpha
_{ij}, \label{U1}%
\end{equation}
where $i$, $j$ represent different site indices, $\alpha_{ij}$ is the gauge
flux and depends on the hopping integral. Here $\sigma_{ij}$ is the Pauli
matrix, whose specific form depends on the gauge vector potential. Therefore,
a tight-binding Hamiltonian of atoms can be written as%
\begin{equation}
\bar{H}=\sum_{\langle i,j\rangle}\left(  t_{ij}\bar{a}_{j}^{\dag}U_{ij}\bar
{a}_{i}+\text{H.c.}\right)  =\bar{H}_{0}+\bar{H}_{\text{SO}} \label{HU}%
\end{equation}
with%
\begin{align}
\bar{H}_{0}  &  =\sum_{\langle i,j\rangle}\left(  t_{ij}\cos\alpha
_{ij}\right)  \bar{a}_{j}^{\dag}\bar{a}_{i}+\text{H.c.}\label{HU1}\\
\bar{H}_{\text{SO}}  &  =i\sum_{\langle i,j\rangle}\left(  t_{ij}\sin
\alpha_{ij}\right)  \bar{a}_{j}^{\dag}\sigma_{ij}\bar{a}_{i}+\text{H.c..}
\label{HU2}%
\end{align}
Here $\langle i,j\rangle$ denotes the nearest-neighbor hoping, $\bar{a}_{i}$
($\bar{a}_{i}^{\dag}$) the creation (annihilation) operator on site $i$. The
first term is the normal Hamiltonian, while the second one is equivalent to SO
coupling. From above equations, we can observe that: (i) By choosing suitable
Pauli matrix $\sigma_{ij}$, one can simulate various SO coupling existing in
actual materials. For our model, we should choose $\sigma_{ij}$=$\sigma_{z}$.
(ii) Through adjusting the gauge flux $\alpha_{ij}$, one can change relative
strength between the two terms. For example, $\alpha_{ij}$=$0$ corresponds no
SO coupling, while $\alpha_{ij}$=$\pi/2$ is equivalent to only the SO coupling
interaction existed in the system. This facilitates us to study the nature
brought by SO coupling. \begin{figure}[ptb]
\begin{center}
\includegraphics[width=0.8\linewidth]{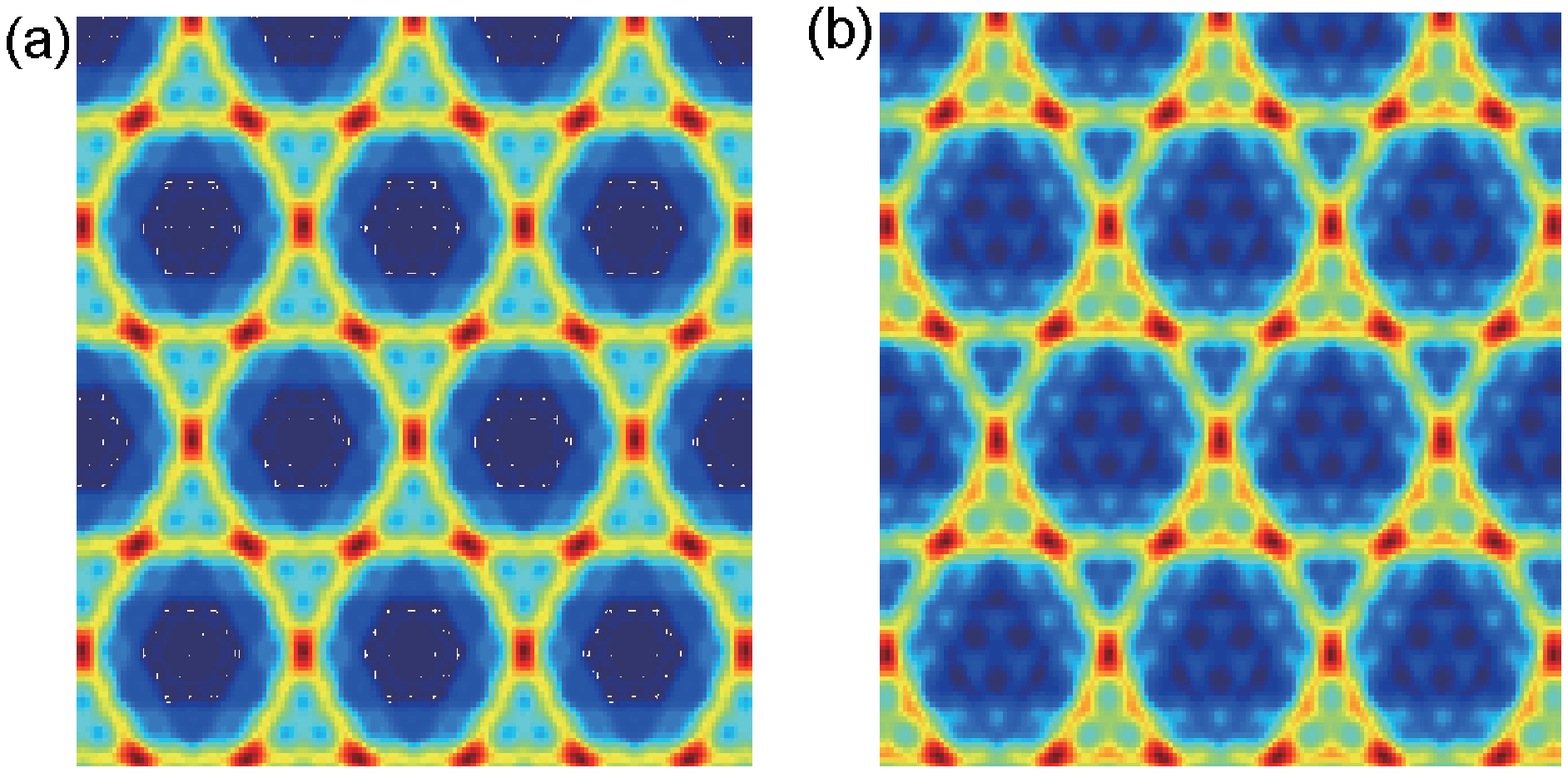}
\end{center}
\caption{(Color online) (a) A uniform Kagom\'{e} lattice for $\varphi$=$\pi$
and $\zeta$=$0$. (b) A trimer Kagom\'{e} lattice for $\varphi$=$\pi$ and
$\zeta$=$1.5$.}%
\end{figure}

We now turn to address the approach to achieve the intrinsic SO coupling
described in Eq.(\ref{HSO}). To this end, we consider a cold atomic system
with each atom having an $\Lambda$-type level configuration (Fig. 5(a)). The
ground states $\left\vert 1\right\rangle $ and $\left\vert 2\right\rangle $
are coupled to the excited state $\left\vert 3\right\rangle $ through
spatially varying standing-wave laser field, with Rabi frequencies $\Omega
_{p}$=$\Omega\sin\theta e^{-iS_{1}}$ and $\Omega_{c}$=$\Omega\cos\theta
e^{-iS_{2}}$, respectively. With rotating-wave approximation, the laser-atom
coupling Hamiltonian is given by
\begin{equation}
\hat{H}_{\text{int}}=-\frac{\hbar}{2}\left(
\begin{array}
[c]{ccc}%
0 & 0 & \Omega_{p}\\
0 & 0 & \Omega_{c}\\
\Omega_{p}^{\ast} & \Omega_{c}^{\ast} & -2\Delta
\end{array}
\right)  \label{HS6}%
\end{equation}
with the eigenstates (the dressing states)$\allowbreak\allowbreak\allowbreak$
\begin{align*}
\left\vert \chi_{1}\right\rangle  &  \text{=}e^{-iS_{1}}\cos\theta\left\vert
1\right\rangle \text{\texttt{-}}e^{-iS_{2}}\sin\theta\left\vert 2\right\rangle
\\
\left\vert \chi_{2}\right\rangle  &  \text{=}\cos\varphi\sin\theta e^{-iS_{1}%
}\left\vert 1\right\rangle \text{+}\cos\varphi\cos\theta e^{-iS_{2}}\left\vert
2\right\rangle \mathtt{-}\sin\varphi\left\vert 3\right\rangle \\
\left\vert \chi_{3}\right\rangle  &  \text{=}\sin\varphi\sin\theta e^{-iS_{1}%
}\left\vert 1\right\rangle \mathtt{+}\sin\varphi\cos\theta e^{-iS_{2}%
}\left\vert 2\right\rangle \text{\texttt{+}}\cos\varphi\left\vert
3\right\rangle
\end{align*}
and eigenvalues $\lambda_{n=1,2,3}$=$0$,$\frac{\hbar}{2}\left(  \allowbreak
\Delta\mathtt{\mp}\sqrt{\Delta^{2}\text{+}\Omega^{2}}\right)  $. Here,
single-photon detuning $\Delta$=$\omega_{3}\mathtt{-}\omega_{1}\mathtt{-}%
\omega_{p}$, with $\omega_{3}$, $\omega_{1}$, $\omega_{p}$ the intrinsic
frequency of atom states $\left\vert 3\right\rangle $, $\left\vert
1\right\rangle $ and laser $\Omega_{p}$, respectively. In the new basis space
$\left\vert \chi\right\rangle $=$\left\{  \left\vert \chi_{1}\right\rangle
\text{, }\left\vert \chi_{2}\right\rangle \text{, }\left\vert \chi
_{3}\right\rangle \right\}  $, the primary atom Hamiltonian $\hat{H}$%
=$\frac{\mathbf{p}^{2}}{2M}$+$\hat{H}_{\text{int}}\left(  \mathbf{r}\right)
$+$\hat{V}\left(  \mathbf{r}\right)  $ can be rewritten as $H$=$\frac{1}%
{2M}\left(  -i\hbar\nabla\mathtt{-}\mathbf{A}\right)  ^{2}$+$V$ with $M$ the
atom mass, $\mathbf{A}$ and $V$ being matrix with matrix element
$\mathbf{A}_{n,m}$=$i\hbar\left\langle \chi_{n}\left(  \mathbf{r}\right)
\mathtt{\mid}\nabla\chi_{m}\left(  \mathbf{r}\right)  \right\rangle $,
$V_{n,m}$=$\lambda_{n}\left(  \mathbf{r}\right)  \delta_{n,m}$+$\left\langle
\chi_{n}\left(  \mathbf{r}\right)  \right\vert \hat{V}\left(  \mathbf{r}%
\right)  \left\vert \chi_{m}\left(  \mathbf{r}\right)  \right\rangle $,
respectively. One can see that in the new basis the atom can be considered as
moving in gauge potential $\mathbf{A}$, which corresponds to an effective
magnetic field $\mathbf{B}_{eff}$=$\left(  \mathbf{\nabla\times A}\right)
\mathtt{-}\frac{i}{\hbar}\left(  \mathbf{A\times A}\right)  $ \cite{Ruse,Zhu3}%
.\begin{figure}[ptb]
\begin{center}
\includegraphics[width=0.8\linewidth]{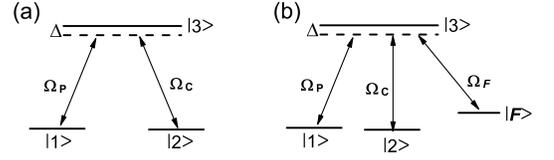}
\end{center}
\caption{(Color online) (a) Illustration of the three-level $\Lambda$-type
atoms coupling with two laser beams with Rabi frequencies $\Omega_{p}$ and
$\Omega_{c}$. (b) Another different hyperfine level $\left\vert F\right\rangle
$ for detection which can be coupled to the excited state $\left\vert
3\right\rangle $ by Rabi frequency $\Omega_{F}$. }%
\end{figure}

We focus on the subspace spanned by the two lower eigenstates $\left\{
\left\vert \chi_{1}\right\rangle \text{, }\left\vert \chi_{2}\right\rangle
\right\}  $, which is redefined by $\left\vert \chi_{\uparrow}\right\rangle
\mathtt{\equiv}\left\vert \chi_{1}\right\rangle $ and $\left\vert
\chi_{\downarrow}\right\rangle \mathtt{\equiv}\left\vert \chi_{2}\right\rangle
$ with the spin language. This gives an effective spin-1/2 system. For the
large detuning ($\Delta\mathtt{\gg}\Omega$) case, both states $\left\vert
\chi_{\uparrow}\right\rangle $ and $\left\vert \chi_{\downarrow}\right\rangle
$ are stable under atomic spontaneous emission from the initial excite state
$\left\vert 3\right\rangle $. Furthermore, we assume the adiabatic condition,
which requires the off-diagonal elements of the matrices $\mathbf{A}$ are
smaller than the eigenenergy differences $\left\vert \lambda_{m}%
\mathtt{-}\lambda_{n}\right\vert $ ($m$, $n$=$1$,$2$,$3$) of the states
$\left\vert \chi_{m}\right\rangle $. Under this adiabatic condition and in the
basis space $\left\{  \left\vert \chi_{\uparrow}\right\rangle \text{,
}\left\vert \chi_{\downarrow}\right\rangle \right\}  $, the gauge potential
$\mathbf{A}$ becomes diagonal and takes the form \cite{Zhu2}%
\begin{equation}
\mathbf{A=}\left(
\begin{array}
[c]{cc}%
\mathbf{A}_{\uparrow} & 0\\
0 & \mathbf{A}_{\downarrow}%
\end{array}
\right)
\end{equation}
with
\[
\mathbf{A}_{\uparrow}=-\mathbf{A}_{\downarrow}=\hbar\left(  \nabla S_{1}%
\cos^{2}\theta\text{+}\nabla S_{2}\sin^{2}\theta\right)  .
\]
Here we neglect the correction to nearest-neighbor tunnelling brought by the
change of potential $V\left(  \mathbf{r}\right)  $ because of the large
detuning approximation.

We consider a specific configuration of the laser beams with two
opposite-travelling standing-wave laser beams \cite{Zhu2,Zhu3}, which take the
Rabi frequencies $\Omega_{p}$=$\Omega\sin\left(  k_{2}y\text{+}\frac{\pi}%
{4}\right)  e^{i\left(  k_{1}x+k_{z}z\right)  }$ and $\Omega_{c}$=$\Omega
\cos\left(  k_{2}y\text{+}\frac{\pi}{4}\right)  e^{-i\left(  k_{1}%
x+k_{z}z\right)  }$. The effective gauge potential is generated as
$\mathbf{A}_{\uparrow}$=$-\mathbf{A}_{\downarrow}$=$\hbar\sin\left(
2k_{2}y\text{+}2\phi_{1}\right)  \left(  k_{1}\mathbf{e}_{x}\text{+}%
k_{z}\mathbf{e}_{z}\right)  $. Here $k_{1}$=$k\sin\theta_{1}\cos\theta_{2}$,
$k_{2}$=$k\cos\theta_{1}$, and $k_{z}$=$k\sin\theta_{1}\sin\theta_{2}$ with
$k$ the wave vector number of laser, $\theta_{1}$ the angle between the wave
vector and $\mathbf{e}_{y}$ axis, $\theta_{2}$ the angle between
$\mathbf{e}_{x}$ axis and the plane consisting of wave vector and
$\mathbf{e}_{y}$ axis. We emphasize that the choice of wave vector $k_{2}$\ of
the laser beams must be a multiple of $\pi/\sqrt{3}$ in order to be
commensurate with the optical lattice. We take $k_{2}$=$\pi/\sqrt{3}$. The
Peierls phase factors for the nearest neighbor hopping in Fig. 1(a) are
$\varphi_{12}^{\alpha}$=$\varphi_{23}^{\alpha}$=$\varphi_{45}^{\alpha}%
$=$\varphi_{56}^{\alpha}$=$\mathbf{-}\alpha\frac{\sqrt{3}k_{1}}{3k_{2}}%
$=$\mathbf{-}\alpha\frac{k_{1}}{\pi}$ and $\varphi_{34}^{\alpha}$%
=$\varphi_{61}^{\alpha}$=$0$ with $\alpha$=$\pm1$ representing the up- and
down-spin. Considering the symmetry of Kagom\'{e} lattice, the vector
potential $\mathbf{A}$ is rotated by $\pm2\pi/3$ to obtain the other two
vector potentials. Therefore, the total effective vector potential and
magnetic field can be written as%
\begin{align}
\mathbf{A}_{eff}^{\alpha}  &  \text{=}\alpha\hbar k_{1}\left[  \left(
\sin\left(  2k_{2}y\right)  \mathtt{-}\cos\left(  k_{2}y\right)  \sin(\sqrt
{3}k_{2}x)\right)  \mathbf{e}_{x}\right. \nonumber\\
&  \left.  +\sqrt{3}\sin\left(  k_{2}y\right)  \cos(\sqrt{3}k_{2}%
x)\mathbf{e}_{y}\right]  , \label{A1}%
\end{align}%
\begin{equation}
\mathbf{B}_{eff}^{\alpha}\text{=}\mathtt{-}\alpha\frac{2\pi\hbar k_{1}}%
{\sqrt{3}}\left[  2\sin(k_{2}y)\sin\left(  \pi x\right)  \text{+}\cos
(2k_{2}y)\right]  \mathbf{e}_{z}. \label{B1}%
\end{equation}
It should be noticed that we have dropped the $\mathbf{e}_{z}$ component in
Eq. (\ref{A1}) because the integral for Peierls phase is only in $x$-$y$
plane. The contours of the magnetic field for up-spin are plotted in Fig.
1(c). However, the total accumulated phases for the nearest-neighbor hopping
along the arrowed directions in Fig. 1(a) are%
\begin{equation}
\varphi_{61}^{\alpha}\text{=}\varphi_{45}^{\alpha}\text{=}\varphi_{34}%
^{\alpha}\text{=}\varphi_{12}^{\alpha}\text{=}\varphi_{23}^{\alpha}%
\text{=}\varphi_{56}^{\alpha}\text{=}\mathtt{-}\alpha\frac{2k\sin\theta
_{1}\cos\theta_{2}}{\pi}\text{=}\alpha\varphi.
\end{equation}
We must retain $\theta_{1}$ to satisfy $k_{2}$=$\pi/\sqrt{3}$. However, we
can\ alter $\varphi$, which controls the relative strength between SO coupling
and ordinary hopping terms, by changing the angle $\theta_{2}$ in the $x$-$z$
plane. So, we can replace\ $t_{ij}\cos\alpha_{ij}\mathtt{\rightarrow}%
t\cos\varphi$ in Eq. (\ref{HU1}) and $t_{ij}\sin\alpha_{ij}\mathtt{\rightarrow
}t\sin\varphi\mathtt{\rightarrow}\lambda_{\text{SO}}$ in Eq. (\ref{HU2}) and
therefore get the intrinsic SO couping model in our cold-atomic Kagom\'{e}
optical lattice.

\section{Detection}

For quantum Hall effect in two-dimensional electronic gas in condensed matter
system, Hall conductivity $\sigma_{xy}$ and Chern number satisfy the relation
$\sigma_{xy}\mathtt{=}\frac{e^{2}}{h}C$ with $h$ the Planck constant and $e$
electronic charge. Thus one can usually detect the Chern number through
measuring Hall conductivity. However, the detection of the spin Chern number
in actual material system is challenge because one can not distinguish the
contributions of the conductivity from up- and down-spin electrons, respectively.

Unlike in electronic system, we will show that a significant advantage of
atomic system is that the spin Chern number can be directly verified by using
the similar method to detect the (mass) Chern number. It has been shown that
the conductivity $\sigma_{xy}$ (Chern number) is related to the atomic density
from the Streda formula $\sigma_{xy}$=$\partial\rho/\partial\mathcal{B}%
\left\vert _{\mu,T}\right.  $when a uniform magnetic field $\mathcal{B}$ is
applied in the system. Thus one can measure the Chern number through the
detection of the density profile, which is a standard detection method used in
atomic system \cite{Umuc,Zhu3}. Since the internal state-dependent image has
also been achieved, the similar method can be straightforward expanded to
measure the spin Chern number. Therefore we may establish the relation between
the spin Chern and atom density, enabling us to detect the OSH phase of the
system via density-profile-measurement technique.\begin{figure}[ptb]
\begin{center}
\includegraphics[width=1.0\linewidth]{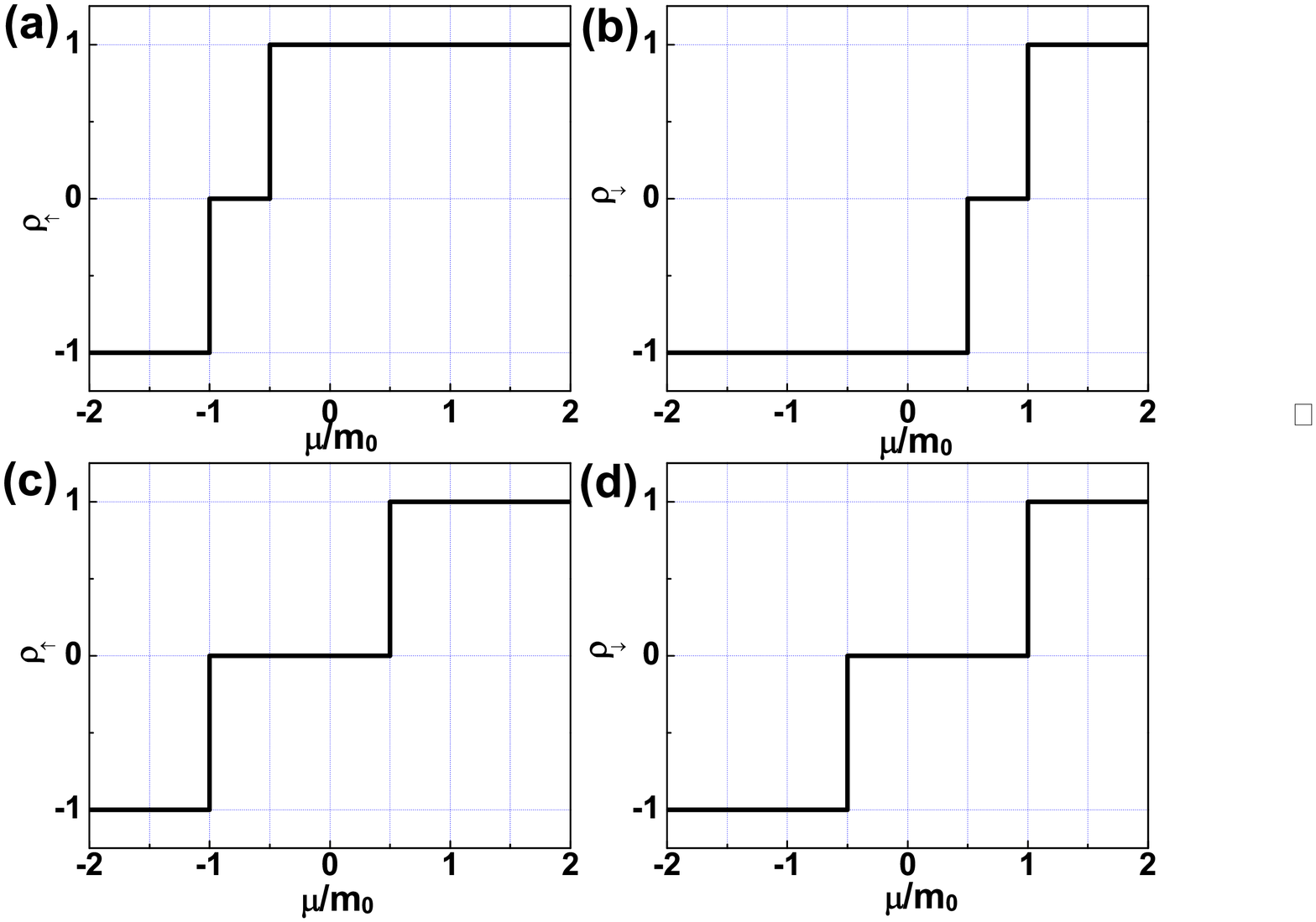}
\end{center}
\caption{(Color online) The spin-atomic density in units $\mathcal{B}/\phi
_{0}$ as a function of the normalized chemical potential $\mu/m_{0}$ with the
definition $\hbar\upsilon_{F}^{2}\left\vert e\mathcal{B}\right\vert
$=$4\left\vert m_{0}\right\vert ^{2}$. (a) and (b) correspond the QSH phase
case with $\sqrt{3}\lambda_{SO}\mathtt{>}3\kappa\mathtt{>}0$; (c) and (d)
correspond the normal phase case with $\sqrt{3}\lambda_{SO}\mathtt{<}%
3\kappa\mathtt{<}0$. $\rho_{\uparrow}$ ($\rho_{\downarrow}$) denotes the
up-(down-)spin-atomic density. }%
\end{figure}

We now introduce how to detect the QSH phase in ultracold-atomic optical
lattice. In cold atomic system, the informations of quantum states are usually
measured from the density profile image. So we will first derive spin-atomic
density from the Dirac Hamiltonian (\ref{HK1}), and then obtain the
information of the Chern number of the system.

The uniform magnetic field can be simulated by rotating the optical lattice at
a constant frequency $\omega$=$e\mathcal{B}/2M$. From the Landau energy levels
obtained at Eq. (\ref{LE}), we get the density \cite{Zhu3} for every Dirac
Hamiltonian (\ref{HK1}) as%
\begin{align}
\rho_{s\alpha}  &  \text{=}\sum_{n=0}^{+\infty}\frac{\text{sgn}\left(
\mu\right)  \mathcal{B}}{2\phi_{0}}\left[  \Theta\left(  \left\vert
\mu\right\vert \mathtt{-}E_{n+1}\right)  \mathtt{+}\Theta\left(  \left\vert
\mu\right\vert \mathtt{-}E_{n}\right)  \right] \nonumber\\
&  \text{+}\frac{\mathcal{B}}{2\phi_{0}}\frac{m_{s\alpha}}{\left\vert
m_{s\alpha}\right\vert }\Theta\left(  \left\vert m_{s\alpha}\right\vert
\mathtt{-}\left\vert \mu\right\vert \right)  , \label{R1}%
\end{align}
where $\Theta$ stands for the unit step function, $\mu$ is the chemical
potential and $\phi_{0}$ the flux quantum. The second term of Eq. (\ref{R1})
is the atom density coming from $n$=$0$ (zero-mode), similarly the discuss of
the Cherm number (Seeing Fig. 2). The calculated spin-atomic density
$\rho_{\alpha}$=$\rho_{+\alpha}$+$\rho_{-\alpha}$ ($\alpha$=$\uparrow
$,$\downarrow$) in unit of $\mathcal{B}/\phi_{0}$ is plotted as a function of
the normalized chemical potential $\mu/m_{0}$ (for $\hbar\upsilon_{F}%
^{2}\left\vert e\mathcal{B}\right\vert $=$4\left\vert m_{0}\right\vert ^{2}$)
in Fig. 6. It is essential that the spatial density profile is uniquely
determined by the function $\rho\left(  \mu/m_{0}\right)  $ in the local
density approximation,\ which is typically well satisfied for trapped
fermions. We focus in the point $\mu$=$0$ where the Dirac mass makes its
central effect on the density or Chern number. When the system is in QSH
phase, as an example, we take the parameters $\sqrt{3}\lambda_{\text{SO}%
}\mathtt{>}3\kappa\mathtt{>}0$, specially assuming $\left\vert m_{+\uparrow
}\right\vert $=$\left\vert m_{-\downarrow}\right\vert $=$0.5\left\vert
m_{-\uparrow}\right\vert $=$0.5\left\vert m_{+\downarrow}\right\vert
$=$0.5m_{0}$, then we have $m_{-\uparrow}\mathtt{>}m_{+\uparrow}%
\mathtt{>}0\mathtt{>}m_{-\downarrow}\mathtt{>}m_{+\downarrow}$. The
spin-atomic density is shown in Fig. 6(a) and (b). It is easy to see that the
up-spin-atomic density $\rho_{\uparrow}$=$\mathcal{B}/\phi_{0}\mathtt{>}0$
(Fig. 6(a)) and the down-one $\rho_{\downarrow}$=$-\mathcal{B}/\phi
_{0}\mathtt{<}0$ (Fig. 6(b)). For the normal phase, we take $3\kappa
\mathtt{>}\sqrt{3}\lambda_{\text{SO}}\mathtt{>}0$ and assume $\left\vert
m_{+\uparrow}\right\vert $=$\left\vert m_{-\downarrow}\right\vert
$=$0.5\left\vert m_{-\uparrow}\right\vert $=$0.5\left\vert m_{+\downarrow
}\right\vert $=$0.5m_{0}$, which means $m_{-\uparrow}\mathtt{>}m_{+\uparrow
}\mathtt{>}0\mathtt{>}m_{-\downarrow}\mathtt{>}m_{+\downarrow}$. We obtain
$\rho_{\uparrow,\downarrow}$=$0$ (Fig. 6(c) and (d)). The spin-atomic density
shows the similar relation comparing to the conclusion of spin Chern number
theroy. This is not surprising because according to the Streda formula and
from Eq. (\ref{R1}), it is easy to obtain the relation between the spin Chern
number and the spin-atomic density as%
\begin{equation}
C_{\alpha}\mathbf{=}\rho_{\alpha}\phi_{0}/\mathcal{B}. \label{Spin-Chern}%
\end{equation}
This formula provides us the approach to measure whether the system is in the
QSH phase. Firstly,\ we measured the spin-atomic density and denoted it as
$\rho_{\uparrow,\downarrow}^{0}$ at $\mu$=$0$ in the absence of $\mathcal{B}$.
Then the optical lattice is rotated to generate the effective uniform magnetic
field $\mathcal{B}$, and the new density of the cold atoms $\rho
_{\uparrow,\downarrow}^{1}$ is measured again. If $\rho_{\uparrow}^{1}%
>\rho_{\uparrow}^{0}$ and $\rho_{\downarrow}^{1}<\rho_{\downarrow}^{0}$,\ the
system is in QSH insulator phase. However, if $\rho_{\uparrow,\downarrow}^{1}%
$=$\rho_{\uparrow,\downarrow}^{0}$, the system is in the normal insulator
phase. Since the density difference is actually quantized in units
$\mathcal{B}/\phi_{0}$,\ the above method could be rather robust.

It is clear from Eq. (\ref{Spin-Chern}) that the total Chern number
$C$=$C_{\uparrow}$+$C_{\downarrow}$=$0$ and thus the direct detection of the
QSH phase is a challenge in an electronic system. However the QSH phase can be
directly verified in the atomic systems since the densities $\rho_{\uparrow}$
and $\rho_{\downarrow}$ can be separately detected \cite{Zhu2}. To
experimentally detect the spin-atomic density, we need first transfer the
dressed state $\left\vert \chi_{\downarrow}\right\rangle $ to a different
hyperfine level $\left\vert F\right\rangle $ which is coupled to the excited
state $\left\vert 3\right\rangle $ by a laser pulse (with a Rabi frequency
$\Omega_{F}$), as seen in Fig. 5(b). This pulse, together with the original
laser beams $\Omega_{p}$ and $\Omega_{c}$, make a Raman transition with an
effective Hamiltonian $H_{R}$=$\left(  \Omega_{F}^{\ast}\Omega/\Delta\right)
\left\vert \chi_{\downarrow}\right\rangle \left\langle F\right\vert $+H.c.
(note that the state $\left\vert \chi_{\uparrow}\right\rangle $ is still
decoupled because of the phase relation between $\Omega_{p}$ and $\Omega_{c}$)
\cite{Zhu2}. Although the form of the state $\left\vert \chi_{\downarrow
}\right\rangle $ is spatially varying, the Rabi frequency $\Omega$ (and thus
also $\Omega_{F}^{\ast}\Omega/\Delta$) is spatially constant. A complete Raman
transition with a $\pi$ pulse will transfer all of the atoms being in the
dressed state $\left\vert \chi_{\downarrow}\right\rangle $ to hyperfine state
$\left\vert F\right\rangle $. After this operation, the initial different
dressed spin states are mapped to different hyperfine levels, and the
populations in different atomic hyperfine levels can be separately imaged with
the known experimental techniques.

\section{Summary}

In summary, we have proposed a model which promises to host the transition
from the QSH insulator phase to the normal insulator phase in the 2D
Kagom\'{e} optical lattice. The model includes two kind of periodic
perturbations, i.e., a nearest-neighbor intrinsic SO coupling and a trimerized
Hamiltonian. The competition between them determines the system's phase. Then
we demonstrate that the lattice SO coupling can be simulated by the
laser-induced-gauge-field method and give the specific laser setting and
parameters to realize the intrinsic SO coupling. Furthermore, we have
established the relation between spin Chern number and spin-atomic density and
then we can detect the spin Chern number through the standard density-profile
technique used in atomic system.

\section{Acknowledgement}

S. L. Zhu was supported in part by NSF of China (No 10974059) and the State
Key Program for Basic Research of China (Nos.2006CB921801 and 2007CB925204).
This work was supported by NSF of China under Grants Nos. 10874235, 10934010,
60978019, and by NKBRSFC under Grants Nos. 2009CB930701, 2010CB922904, 2011CB921500.

\end{document}